# Black hole in a waveguide: Hawking radiation or self-phase modulation?

Igor I. Smolyaninov

*Department of Electrical and Computer Engineering, University of Maryland, College Park, MD 20742, USA*

**Recently it was suggested that Hawking radiation may be observed in a nonlinear electromagnetic waveguide upon propagation of an optical pulse. We show that the spectral characteristics of the Hawking effect in such a waveguide are indistinguishable from the well-known effect of frequency broadening of an optical pulse due to self-phase modulation. Furthermore, we derive an estimate on the critical optical power at which Hawking effect is dominated by the self-phase modulation. It appears that optical experiments reported so far are clearly dominated by self-phase modulation.**

In 1974 Hawking demonstrated that black holes can evaporate by emission of thermal radiation [1]. A closely related effect had been introduced a few years later by Unruh. He showed that for an accelerating observer vacuum should look like a bath of thermal radiation with temperature $T_{UH}$ defined as

$$k_B T_{UH} = \frac{\hbar a}{2\pi c} \qquad (1)$$

where *a* is the acceleration of the observer [2]. The Hawking temperature may be obtained from eq.(1) by substitution of the free fall acceleration near the black hole horizon into eq.(1). While well-established theoretically, these effects are believed to be very difficult to observe in the experiment: an observer accelerating at $a=g=9.8$m/s$^2$ should see vacuum temperature of only $4 \times 10^{-20}$K. Over the past years quite a few proposals were introduced on how to emulate and/or observe the Unruh-Hawking effect. A recent review of these proposals, which involve various acoustic, optical, accelerator physics, and solid state physics experimental situations can be found in [3]. Very recently it was suggested that the Hawking radiation may potentially be observed in optical waveguides [4]. In other recent proposals it was suggested that the Unruh-Hawking radiation may have already been detected in the experiments with surface plasmon polaritons [5] which propagate over a curved metal surface [6], and that the Unruh effect may be emulated in tapered optical waveguides [7]. However, despite large number of different proposals and even claimed experimental observations [8], so far there was no undisputed report on experimental observation of the Hawking radiation in any such analog system (see for example comment [9] on ref.[8]).

In this paper we demonstrate that in the recently proposed model of Hawking effect in an electromagnetic waveguide [4], the spectral characteristics of Hawking radiation are indistinguishable from the well-known effect of frequency broadening of an optical pulse due to self-phase modulation. Furthermore, we derive an estimate on the critical optical power at which Hawking effect is dominated by the self-phase modulation. It appears that optical experiments reported so far are clearly dominated by self-phase modulation.

Following ref.[4] and the closely related ref.[10] let us consider an initially (almost) monochromatic long optical pulse of frequency $\omega_0$, which propagates along an optical fiber as shown in Fig.1. At large enough intensity $I$ of the pulse the refractive index of the fiber $n$ is modified by the propagating pulse due to the Kerr nonlinearity:

$$n = n_0 + n_2 I = n_0 + \frac{\chi^{(3)}}{2n_0} I, \qquad (2)$$

where $\chi^{(3)}$ is the third order nonlinear susceptibility. For the sake of simplicity let us assume that the $n_2$ coefficient does not exhibit any dependence on frequency. On the other hand, let us assume that the waveguide exhibits anomalous dispersion: $dn_0/d\omega < 0$ in the wide enough frequency range around $\omega_0$. This means that at low frequencies ($\omega < \omega_0$) the speed of photons in the waveguide is smaller than the speed of the propagating optical pulse. In general, the two conditions $n_2 > 0$ and $dn_0/d\omega < 0$ correspond to the right conditions for the formation of an optical soliton in the fiber. However, for the sake of simplicity let us not be concerned about the stability of the propagating optical pulse. The situation shown in Fig.1 in the reference frame co-moving with the optical pulse replicates the picture of two horizons for the low-frequency photons, which is similar to the one considered in Fig.2 of ref.[10]: the front edge of the optical pulse behaves as a black hole horizon for low frequency photons in the fiber, while the trailing edge of the pulse behaves as a white hole horizon. The low frequency photons (if any) in front of the front edge of the pulse cannot escape it, because their speed is smaller than the speed of the pulse. On the other hand, the low frequency photons behind the trailing edge of the pulse cannot reach the trailing edge. This qualitative picture of the toy "black hole" and "white hole" horizons in an optical fiber is intuitively clear. The mathematical justification of this picture in terms of the effective metric, effective surface gravity, etc. which is experienced by the low frequency photons can be found in refs.[4,10]. The resulting expression for the temperature of Hawking radiation may be found in refs.[4,7,10,11]:

$$k_B T_{UH} = \frac{\hbar}{2\pi}\left|\frac{dc^*}{dx}\right| , \qquad (3)$$

where the gradient of the speed of light $c^*$ in the fiber is taken at the emulated black hole horizon. The corresponding frequency band $\Delta\omega_{UH}$ of the effective Hawking radiation (which may be defined as $\hbar\Delta\omega_{UH}=k_B T_{UH}$) is

$$\Delta\omega_{UH} = \frac{1}{2\pi}\left|\frac{dc^*}{dx}\right| \qquad (4)$$

Using eq.(2) we can re-write eq.(4) as

$$\Delta\omega_{UH} = \frac{k_B T_{UH}}{\hbar} = \frac{c}{2\pi n^2}\left|\frac{dn}{dx}\right| = \frac{1}{2\pi n}\left|\frac{dn}{dt}\right| = \frac{n_2}{2\pi n}\left|\frac{dI}{dt}\right| , \qquad (5)$$

where we have replaced the refractive index gradient by the time derivative of the refractive index (this is a valid replacement since similar to ref.[4] the local speed of light variations due to the propagation of the pulse through the waveguide depend on $(x+c^*t)$).

We may also calculate the emissive power per unit surface area of the Hawking radiation in the waveguide using the Stefan-Boltzmann law. In the co-moving reference frame the emissive power is

$$P_{UH} = \sigma T_{UH}^4 = \frac{\sigma\hbar^4 n_2^4}{16\pi^4 k_B^4 n^4}\left|\frac{dI}{dt}\right|^4 = \frac{\hbar n_2^4}{960\pi^2 c^2 n^4}\left|\frac{dI}{dt}\right|^4 , \qquad (6)$$

where $\sigma$ is the Stefan-Boltzmann constant. The Hawking radiation emissive power is supposed to be proportional to $n_2^4 |dI/dt|^4$. In the laboratory frame at the fiber output the observed temperature of a moving black body is shifted due to Doppler effect as

$$T_{UH}^* = T_{UH} \sqrt{\frac{c^*+c}{c^*-c}} = T_{UH} \sqrt{\frac{n+1}{n-1}} \quad , \tag{7}$$

Therefore, the spectral energy density at the fiber output is given by the Planck law as

$$I_{UH} = \frac{\hbar \omega^3}{16\pi^2 c^2 \left( \exp\left( \frac{\hbar \omega}{k_B T_{UH}^*} \right) - 1 \right)} \tag{8}$$

Let us confront these results with the well-known results for the effect of self-phase modulation in a nonlinear waveguide. Unfortunately, the frequency band $\Delta\omega_{UH}$ of the effective Hawking radiation in eq.(5) appears to be identical to the well-known effect of spectral broadening of an optical pulse due to self-phase modulation. Indeed, the effect of self-phase modulation is usually derived by considering local phase variations across the optical pulse due to intensity-dependent refractive index described by eq.(2). The local shift in frequency $\Delta\omega(x)$ at the given location $x$ in the pulse in the reference frame co-moving with the pulse is obtained as

$$\omega_0 + \Delta\omega(x) = \frac{d\phi(x)}{dt} = \frac{d}{dt}(\omega_0 t - kx) \quad , \tag{9}$$

so that

$$\Delta\omega(x) = -\left(\frac{2\pi x}{\lambda_0}\right)\frac{dn}{dt} = \left(\frac{2\pi x}{\lambda_0}\right) n_2 \left|\frac{dI}{dt}\right| \quad , \tag{10}$$

where $\lambda_0$ is the wavelength of light in free space. Thus, spectral width of the pulse is broadened near the leading and the trailing edges of the pulse where $dI/dt$ is nonzero. The constant factor $\left(\frac{2\pi x}{\lambda_0}\right)$ of the order of unity is defined by the spatial shape of the pulse (shown in Fig.1). The spectral broadening of the pulse obtained via well-established nonlinear optical effect of self-phase modulation appears to be the same as the one (eq.(5)) obtained based on the expression for the effective Hawking temperature, which is derived in ref. [4]. Both expressions are defined by $n_2\left|\frac{dI}{dt}\right|$ multiplied by a factor of the order of unity. Moreover, since fiber considered in this experimental geometry exhibits Kerr nonlinearity given by eq.(2), the nonlinear optical interaction of original $\omega_0$ quanta and the $\omega_0+\Delta\omega$ quanta produced at the leading and trailing edges of the pulse will lead to difference frequency generation of photons at $\omega \sim \Delta\omega$ frequencies. This difference frequency generation will lead to observation of Hawking-like radiation spectrum at low frequencies, which is indistinguishable from the effect described by eqs.(3-5). These general arguments may be confirmed by calculations of the pulse evolution due to self-phase modulation using formalism developed in [12].

The evolution of optical pulse due to Kerr nonlinearity (given by eq.(2)) may be determined by solving the nonlinear Schrödinger equation [12] for the electric field $E(x,t)$ in the pulse

$$\left(\frac{d}{dx}+\frac{n_0}{c}\frac{d}{dt}\right)E = -n_2 E^2 \frac{dE}{cdt} \quad , \tag{11}$$

where *x* is the propagation direction. As demonstrated in [12], the solution of eq.(11) may be found using the method of successive approximations:

$$E = E^{(0)} + n_2 E^{(1)} + n_2^2 E^{(2)} + \dots , \qquad (12)$$

The first order correction to $E^{(0)}$ gives rise to generation of $\omega + \Delta\omega$ frequencies at the leading and trailing edges of the pulse:

$$E^{(1)} = -\frac{n_2 x}{c} E^{(0)2} \frac{dE^{(0)}}{dt} \qquad (13)$$

(as demonstrated by eqs.(9,10)). The second order correction to $E^{(0)}$ gives rise to generation of $\Delta\omega$ frequencies at the leading and trailing edges:

$$E^{(2)} = -\frac{n_2 x}{c} \left(E^{(0)} + E^{(1)}\right)^2 \frac{d(E^{(0)} + E^{(1)})}{dt} \qquad (14)$$

The field at the $\Delta\omega$ frequencies may be estimated as

$$E^{(2)}_{\Delta\omega} \propto -\frac{n_2 x}{2c} E^{(1)} \frac{dI}{dt} = \frac{n_2^2 x^2}{4c^2} I^{1/2} \left(\frac{dI}{dt}\right)^2 \qquad (15)$$

leading to the final estimate of the field intensity at $\Delta\omega$:

$$I^{(2)}_{\Delta\omega} \propto \frac{x^4}{16c^4} In_2^4 \left(\frac{dI}{dt}\right)^4 \qquad (16)$$

Thus, in agreement with general arguments described above, the second order corrections to the pulse shape due to self-phase modulation lead to generation of Hawking-like $\Delta\omega$ frequencies, and the field intensity at these frequencies is proportional to $n_2^4 |dI/dt|^4$, which once again looks identical to the result obtained for the Hawking radiation. However, comparison of eqs.(6) and (16) clearly indicates that self-phase modulation dominates Hawking radiation in any realistic nonlinear optics geometry. Assuming pulse width and the waveguide core dimensions of the order of light wavelength $\lambda_0$, the critical power at which self-phase modulation starts to dominate Hawking radiation may be estimated as

$$P_c = \frac{\hbar c^2}{60\pi^2 \lambda_0^4} \sim 26 \text{ W/cm}^2 \qquad (17)$$

in the visible frequency range. Given the fiber core radius ~ 1 μm, this estimate corresponds to ~0.26 μJ of optical energy in the pulse. These numbers, are much smaller than the power levels used in [8]. Moreover, if the measured value of $n_2$=2.8x10$^{-16}$ cm$^2$/W of the fused silica [8] is taken into account, the critical change in refractive index $\Delta n_c$ may be estimated as 7x10$^{-15}$. Such a low critical value for the Hawking effect observation does not look encouraging from the experimental standpoint.

The effect of self-phase modulation was seen in numerous experiments. The most striking examples of this effect are described in refs.[13,14] in which the effective Hawking temperature given by eq.(3) is also very large, and a broad white light supercontinuum is generated while intense femtosecond optical pulses propagate through optical fibers exhibiting anomalous dispersion. However, in all the nonlinear optical experiments reported so far the Hawking effect is clearly dominated by the self-phase modulation, while spectral characteristics of the Hawking effect are indistinguishable from the second order effects of self-phase modulation. Given the really low critical threshold power (eq.(17)) at which Hawking radiation may potentially be observed, the perspective of experimental observation of the Hawking effect in a waveguide does not look encouraging.

**Figure Captions**

**Figure 1.** Light intensity and the speed of light in an optical fiber in the reference frame co-moving with the intense optical pulse. Low frequency photons cannot escape the leading edge of the pulse, which constitutes an effective black hole horizon for low frequencies. On the other hand, the low frequency photons behind the trailing edge of the pulse cannot reach the trailing edge, making it an effective white hole.

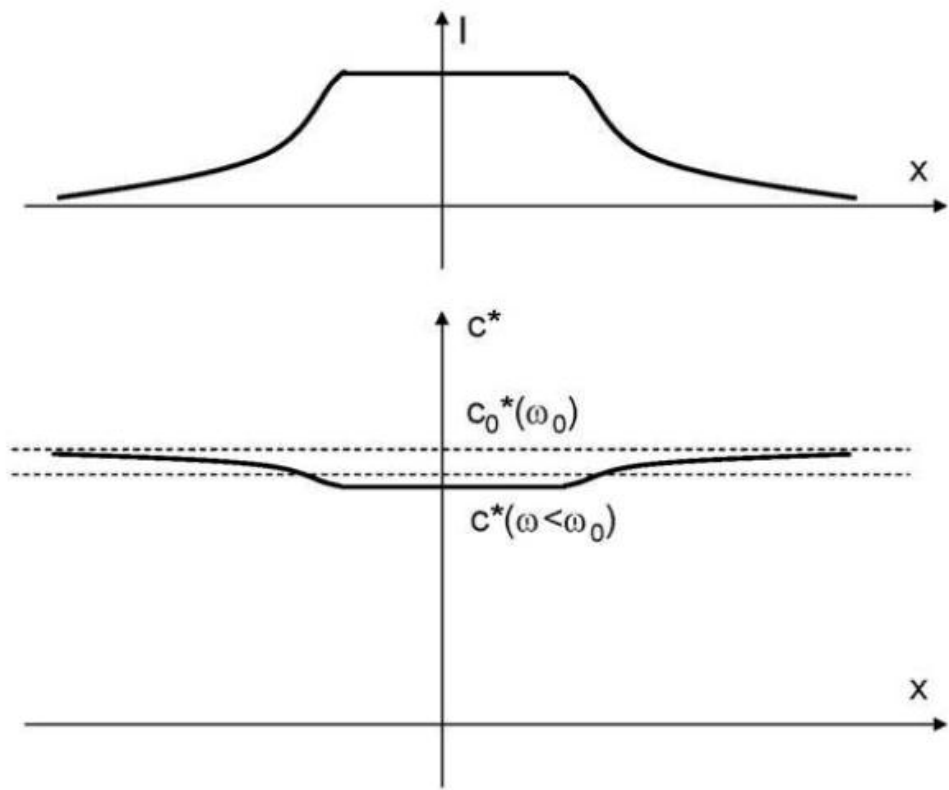

Fig. 1